%% file: main.tex
\documentclass[a4paper,11pt]{article}
\pdfoutput=1 

\usepackage{jcappub} 
\usepackage[]{natbib}
\usepackage[T1]{fontenc} 
\usepackage{graphicx}
\usepackage{subcaption}
\usepackage{enumitem}
\usepackage{xcolor}
\usepackage{soul}
\usepackage{float}
\begin{document}

\title{Constraints on Neutrino Emission from Nearby Galaxies Using the  2MASS Redshift Survey and IceCube}
\collaboration{IceCube Collaboration}
\emailAdd{analysis@icecube.wisc.edu}

\input{authors_jcap.tex}
\abstract{
The distribution of galaxies within the local universe is characterized by anisotropic features. Observatories searching for the production sites of astrophysical neutrinos can take advantage of these features to establish  directional correlations between a neutrino dataset and overdensities in the galaxy distribution in the sky. The results of two correlation searches between a seven-year time-integrated neutrino dataset from the IceCube Neutrino Observatory, and the 2MASS Redshift Survey (2MRS) catalog are presented here. The first analysis searches for neutrinos produced via interactions between diffuse intergalactic Ultra-High Energy Cosmic Rays (UHECRs) and the matter contained within galaxies. The second analysis searches for low-luminosity sources within the local universe, which would produce subthreshold multiplets in the IceCube dataset that directionally correlate with galaxy distribution. No significant correlations were observed in either analyses. Constraints are presented on the flux of neutrinos originating within the local universe through diffuse intergalactic UHECR interactions, as well as on the density of standard candle sources of neutrinos at low luminosities. 
}

\maketitle
\flushbottom

\section{Introduction}
\label{sec:intro}

 The IceCube Neutrino Observatory is a cubic-kilometer of instrumented glacial ice buried from 1.5 km to 2.5 km below the surface of the geographic South Pole.  It contains an array of 5160 PMTs to detect the Cherenkov radiation from charged particles passing through the detector and is designed to search for neutrinos from astrophysical sources \cite{Aartsen:2016nxy}. Astrophysical neutrinos can be identified in two ways: by searching for neutrinos with interaction vertices inside the detector, or by correlating the arrival directions of well-reconstructed through-going events with astrophysical sources or with each other \cite{Aartsen:2014gkd, Aartsen:2016xlq}. Both types of searches have produced compelling detections of astrophysical neutrinos.  IceCube ushered in the advent of neutrino astronomy with the statistically significant detection of a diffuse flux of astrophysical neutrinos using starting events.  More recently, IceCube reported evidence of the first astrophysical source of high-energy neutrinos \cite{IceCube:2018dnn,IceCube:2018historicsearch} using a larger dataset of through-going events. However, the discovery of individual neutrino sources via a $5\sigma$ detection in the accumulated IceCube dataset remains elusive \cite{IceCube:7yrPS}.
\\

 Anisotropy is observed in the nearby universe with large superclusters, such as the supergalactic plane, creating overdense regions. Since a predominantly Galactic origin for the high-energy neutrino flux has been disfavoured by IceCube \cite{IceCube:galactic,Albert:2018vxw}, the local universe becomes the next nearest location for possible neutrino production \footnote{While general assumptions about the redshift evolution of neutrino sources do not necessarily favor sources in the local universe \cite{Ahlers:2014ioa}, a search for correlations with nearby objects is motivated by 1/$r^2$ losses expected in the observed neutrino flux.}. A potential link between the diffuse neutrino flux and the local universe could be revealed by likelihood-based methods mapping neutrino locations to astrophysical objects or structures in the sky, which have been used in other studies \cite{icecube_fermiBLAC}.\\

In this paper two searches for correlations between neutrinos and local large-scale structures are presented, each of them targeting a different mechanism responsible for the neutrino emission. The first, which will be referred to as the \textit{template analysis}, searches for  UHECRs interacting hadronically with matter.  To first order, the flux of UHECRs is isotropic, but the distribution of large-scale structure in the local universe is not.  Therefore, directional correlation will be observed between the overdensities of the local universe and astrophysical neutrinos. The second analysis, referred to as the \textit{multiplet analysis}, tests for the existence of low-luminosity neutrino sources that occur more frequently in dense regions of the local universe. Such sources will produce spatial clusters of neutrinos (referred to as \textit{multiplets} in this paper) that directionally correlate with galaxy density \cite{Mertsch:2016hcd}. Section \ref{sec:dataset} describes the source catalog that is used in both analyses as a tracer for the matter distribution in the local universe. The high statistic neutrino dataset used in both searches is also described. The details of the two analyses are described in section \ref{sec:method} and their results are presented in section \ref{sec:results}.\\

\section{Datasets}
\label{sec:dataset}

\begin{figure}[ht]
\centering
\includegraphics[width=.6\textwidth]{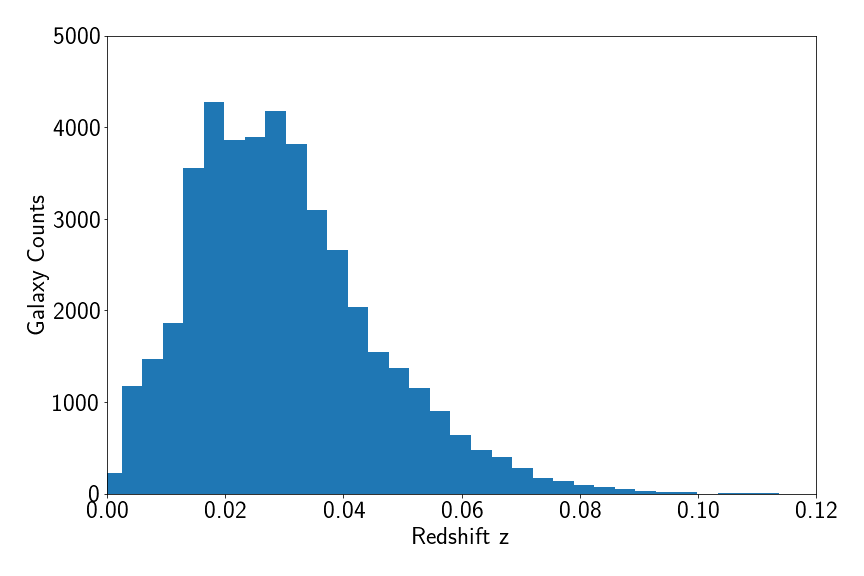} 
\caption{Distribution of galaxies in 2MRS by measured redshift. Magnitude limit due to instrumental sensitivity becomes dominant beyond a redshift of $z=0.03$.}
\label{fig:2MRS}
\end{figure}

To study correlations between neutrino arrival directions and the nearby matter distribution, it is necessary to select an appropriate tracer of large-scale structure that maximizes sky coverage and minimizes observational biases. For this study the 2MASS Redshift Survey (2MRS) galaxy catalog is used. 2MRS is an all-sky catalog of galaxies with associated redshift measurements, and is the most extensive and unbiased available survey up to redshifts of $z <  0.03$ \cite{2mrs}.  The survey was performed at infrared wavelengths, and contains the position, redshift and $K_s$ magnitude of the galaxies of the local universe up to a redshift of $z \approx 0.10$.  The distribution of redshift for all galaxies in the catalog is shown in Figure \ref{fig:2MRS}. The anisotropic features in this distribution have been used in other correlation searches, namely by the Auger collaboration \cite{auger_2mrs, auger_uhecr}. A magnitude limit, imposed by instrument sensitivity, is important above $z=0.03$, but the catalog can be considered complete below this redshift. The uses of the catalog for sources at $z>0.03$ is discussed in section \ref{sec:method}.\\

The dataset used for the \textit{template} and \textit{multiplet} analyses consists of 7 years of candidate muon neutrinos and atmospheric muons which produce track-like events in the IceCube detector. After processing and selection cuts, the dataset consists of 422,791 upgoing events from the Northern Hemisphere and 289,078 downgoing events from the Southern Hemisphere detected over 2,431 days of detector exposure in the years of 2008-2015.\\

The 7-yr event sample has been used for other neutrino searches, and details of the event selection are outlined in \cite{IceCube:7yrPS}. In brief, separate quality cuts are applied to events from the Northern and Southern Hemispheres because the dominant sources of background differ as a function of zenith. In the southern (downgoing) direction, a large background of atmospheric muons from cosmic rays requires restrictive cuts on event energy. Muons from the northern sky (upgoing direction) are blocked by the Earth, so less restrictive energy cuts are applied, leaving a primary irreducible background of atmospheric neutrinos from the Northern Hemisphere. There is an all-sky astrophysical neutrino flux from both hemispheres that is 3 to 4 orders of magnitude lower than the respective backgrounds in each part of the sky.  The event selection is optimized for correlations with point-like neutrino sources, so the dataset will be referred to as the 7-yr PS sample. After all quality selections are made, 90\% of events fall within the energy range of 400\,GeV to 225\,TeV, and the median angular resolution of the sample is better than $1^\circ$ above 1 TeV. Further details are available in \cite{IceCube:7yrPS}.\\

\section{Analysis Methods}
\label{sec:method}

\subsection{Template Analysis}
The goal of the \textit{template analysis} is to find statistical correlations between directions of astrophysical neutrinos and local galaxy density. A Test Statistic (TS) is defined to measure the similarity between the spatial distribution of neutrinos observed in IceCube with a spatial template of galaxy densities obtained from 2MRS.  The TS gives higher weight to high-energy neutrinos, since the astrophysical neutrinos are modeled by an $E^{-2}$ energy spectrum and the atmospheric muon and neutrino backgrounds follow an $E^{-3.7}$ spectrum \cite{Gaisser:2002jj}.  The TS also includes different spatial probability density functions (PDFs) for the signal and background hypotheses. The signal PDF comes from the 2MRS catalog, weighted by the sensitivity of IceCube as a function of declination; while the background PDF assumes a uniform distribution of atmospheric muons and neutrinos in right ascension (RA), weighted by the detector sensitivity.  The statistical significance of the correlation found in the data is quantified as a p-value by comparing the experimental values of TS with a distribution of TS obtained from datasets with randomized RA coordinates.

\subsubsection{Likelihood function and Test Statistic} \label{template:LLH}
The \textit{template analysis} is based on a likelihood function \cite{Braun:2008} given by a product over the $N$ total events in the dataset:\begin{equation} \label{eq:1}
L(n_s) = \prod_{i=1}^{N}\left( \frac{n_s}{N} S_i(\mathbf{x}_i, \sigma_i, E_i) + (1-\frac{n_s}{N})B_i(\mathbf{x}_i, E_i) \right),
\end{equation} where $n_s$ is the number of signal events. $S_i$ is the likelihood of event $i$ contributing to the signal which is a function of particle event direction $\mathbf{x}_i$, energy $E_i$ and angular uncertainty $\sigma_i$. $B_i$ is the likelihood of event $i$ contributing to the background, which is a function of particle event direction and energy only.  $S_i$ and $B_i$ are the products of spatial and energy PDFs specific to the signal and background hypotheses, respectively. The likelihood is a function of ${n_s}$, which is free to vary between 0 and $N$. The entire dataset is fit for the most likely number of ``signal'' neutrinos correlating with large-scale structure, which is denoted $\hat{n}_s$.  The test statistic is defined as:\\
\begin{equation} \label{eq:2}
\mathrm{TS} = -2\ln{\left(\frac{L(n_s=0)}{L(\hat{n}_s)}\right)} \\
\end{equation}
where $L(n_s=0)$ is the likelihood for the hypothesis corresponding to no correlation, and $L(\hat{n_s})$ is the likelihood for the best-fit $\hat{n}_s$.

\subsubsection{Signal Hypothesis}\label{template:PDF}

\begin{figure}[ht]
\includegraphics[width=.88\textwidth]{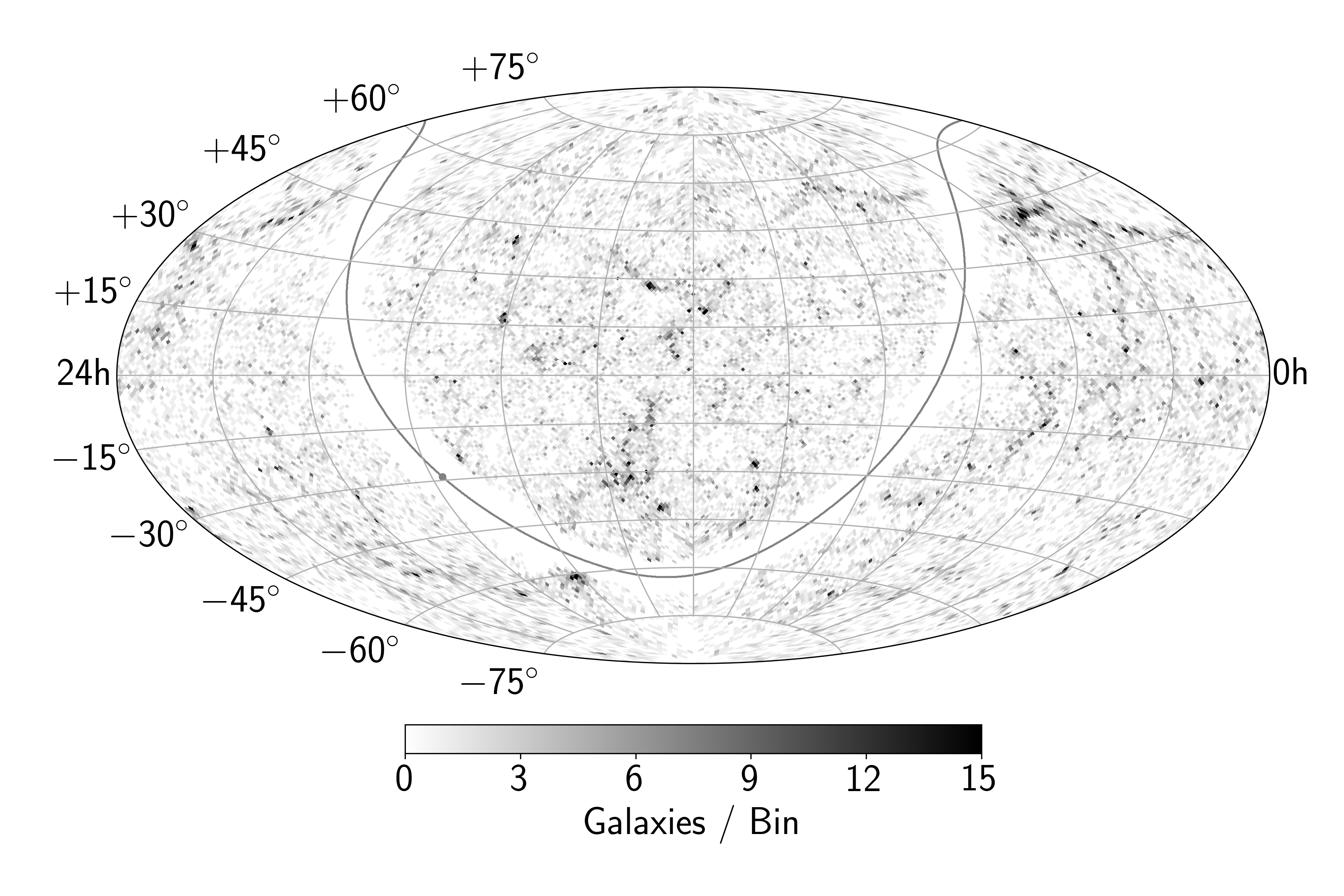}  \\
\includegraphics[width=.88\textwidth]{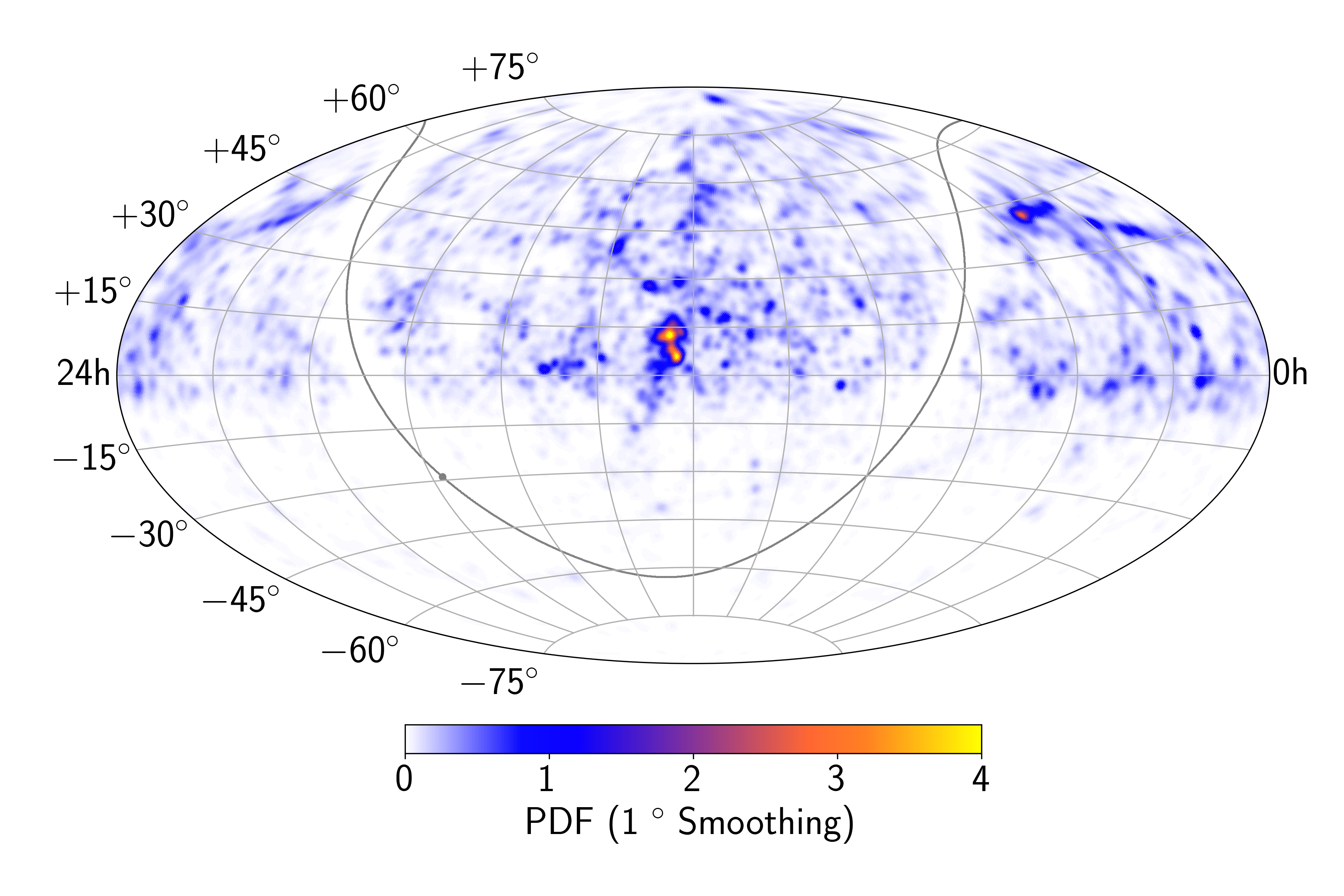} \\
\caption{\textit{Template analysis} template with galaxies weighted by redshift distance shown here in equatorial coordinates.  The PDF is constructed by taking maps of galaxy density (Top), weighting and convolving with the detector acceptance for an $E^{-2}$ spectrum and convolving with a Gaussian here shown for events with uncertainty = 1$^\circ$, resulting in final spatial signal PDF (Bottom), dominated by the local Virgo Supercluster.}
\label{fig:PDF}
\end{figure} 

The \textit{template analysis} uses spatial PDFs, or ``templates,'' based on the 2MRS catalog.  The templates are based on three different galaxy weighting schemes:
\\
\begin{itemize}[noitemsep]
\item All galaxies with $z$ \textless 0.03 equally weighted 
\item All galaxies weighted by redshift  
\item All galaxies equally weighted
\end{itemize}

The weighting schemes are chosen to probe specific correlation scenarios. The template with a cut at $z$=0.03 corresponds to a complete but locally restrictive catalog.  The redshift-weighted template tests if the neutrino flux is proportional to the inverse square of the distance\footnote{Note that the effect of peculiar motion on the redshift-distance calculation is not included in the weights.}. Using the full catalog with equal weights favors correlations with the most luminous galaxies, since 2MRS is magnitude-limited beyond $z=0.03$. Thus, this scheme assumes that the galaxies most likely to be neutrino sources are the brightest ones.\\

Templates are constructed by first creating density maps of the 2MRS galaxies using the HEALPix pixelization of the unit sphere \cite{HEALPix}. The maps use equal-area pixels with a solid angle of 0.2 square degrees. Maps are then weighted according to the three scenarios described above, convolved with the IceCube detector acceptance function (calculated for an $E^{-2}$ spectrum), and the maps are smeared using a Gaussian of width $\sigma$ representing the uncertainty in the arrival direction of each neutrino event.  By building templates based on each event, the unique uncertainty in the angular reconstruction of each neutrino candidate is used in \ref{eq:1} when computing the correlation with a 2MRS template.  In order to optimize computation efforts, the maps are pre-built for each $\sigma$ in the range 0$^\circ$ to 4$^\circ$ with steps of 0.1$^\circ$. An example of a final template is shown in Figure \ref{fig:PDF}.\\

\subsection{Multiplet Analysis}

The multiplet analysis is sensitive to a directional excess of neutrino \textit{multiplets} that correlate with the local galaxy density. This analysis uses a previously published calculation of neutrino spatial clustering performed on the 7-yr PS sample \cite{IceCube:7yrPS}. The study searched for statistically significant clusters of neutrino events in the data, and produced a high-spatial resolution skymap describing the statistical significance of all neutrino clusters identified in the data, along with their associated $p$-values, the best-fit number of signal neutrinos $n_s$ at each location in the sky, and the best-fit spectral index $\gamma$ at each location. It has been suggested that the local maxima in the significance map could originate from an underlying population of low-luminosity sources in the local universe \cite{Mertsch:2016hcd}. These local maxima, called \textit{multiplets}, are the inputs to the analysis.\\

The \textit{hotspot} tool from HEALPix \cite{HEALPix} is used to locate all local maxima within the significance map. A cleaning algorithm is used to avoid double-counting a maximum, since the significance map pixelization is finer than the detector's angular resolution. All pixels within a distance of 1.5$^{\circ}$ from the most significant local maximum are removed, and the procedure is recursively applied to the next most significant maximum remaining on the cleaned map. The distance cut is motivated by a calculation in \cite{PStechnique} based on the angular resolution of neutrino events in the used sample.\\

\begin{figure}[ht]
    \begin{subfigure}[b]{0.88\textwidth}
        \includegraphics[width=\textwidth]{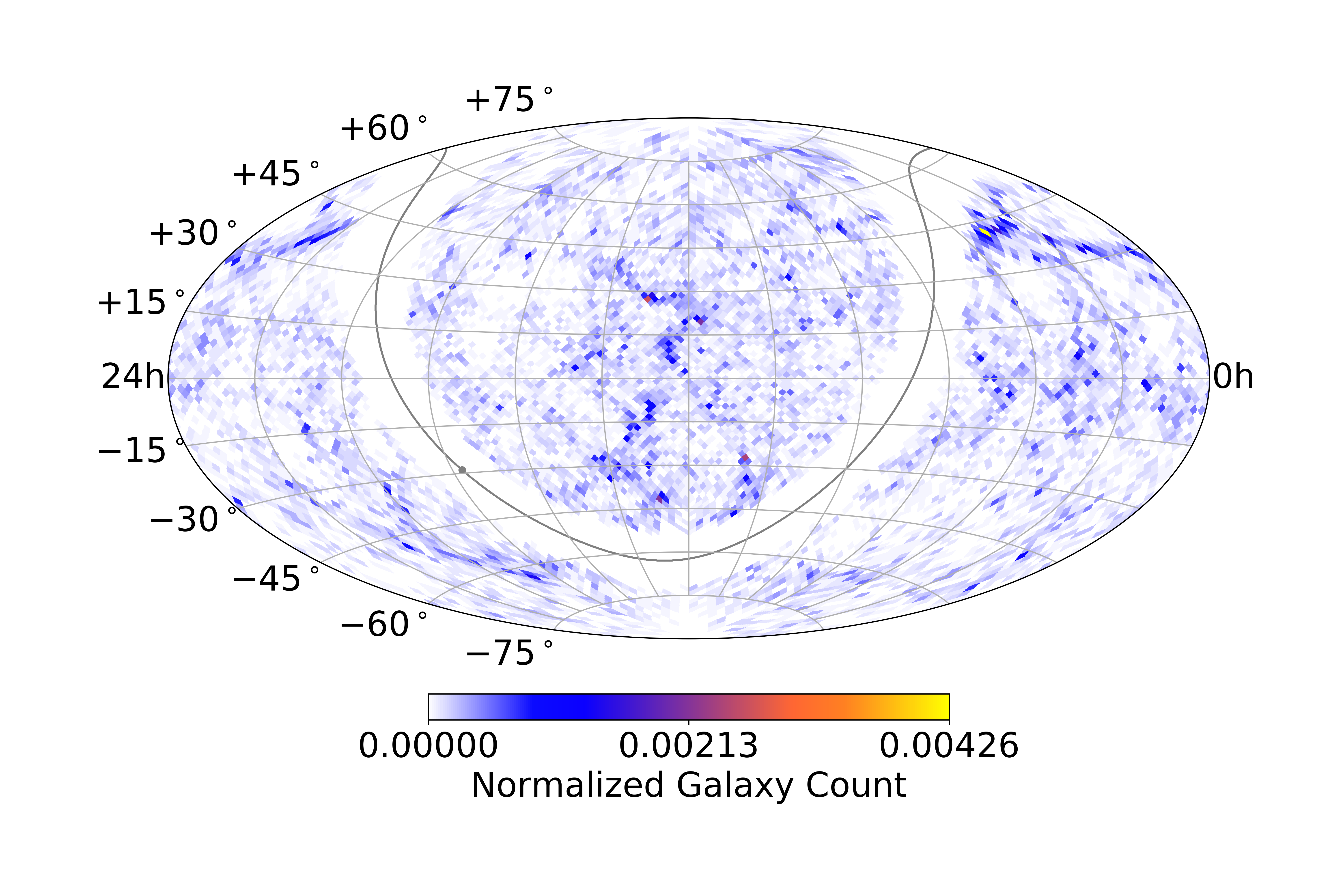}
        \caption{ }
        \label{multiplet_signal}
    \end{subfigure}
    
    \begin{subfigure}[b]{0.88\textwidth}
        \includegraphics[width=\textwidth]{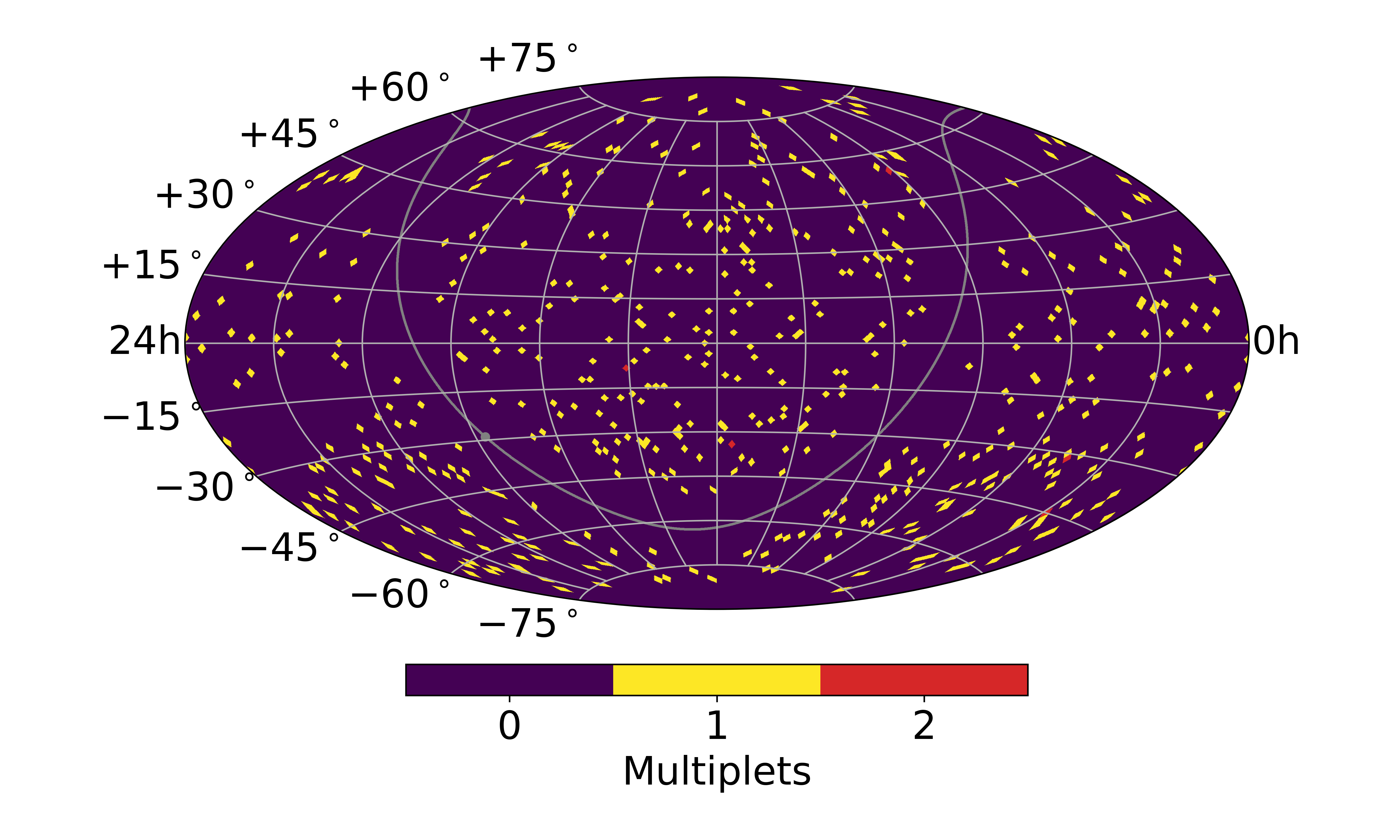}
        \caption{ }
        \label{multiplet_multiplets}
    \end{subfigure}
    
    \caption{a)  Normalized distribution of galaxies up to a redshift of 0.03, taken from the 2MASS Redshift Survey catalog \cite{2mrs}.  b) Location of the selected subset of multiplets. Each yellow tile represents the location of a local maximum from the  7yr-PS significance map which satisfies the selection criteria of $n_s\geq 2.0$ and $\gamma \geq -2.75$. Red tiles denote the five instances in which two multiplets fall into the same pixel. These are therefore counted twice in the likelihood calculation}
    \label{multiplets_figs}
\end{figure}

Local maxima are only kept as multiplets if their local pre-trial significance is greater than 2.0$\sigma$. Furthermore, the best-fit number of signal neutrinos at that location (the variable $n_{s}$ in the 7yr-PS likelihood) is required to be greater or equal to 2. Finally, an additional cut on the local energy spectrum (the local spectral index $\gamma\geq-2.75$) is applied to filter out multiplets which are most likely coming from atmospheric neutrino background events. The choice of cut thresholds is based on an optimization study where the figure of merit took into account sample statistics and sensitivity. The final multiplet sample is shown in Fig. \ref{multiplet_multiplets}.\\

Since the 2MRS catalog does not cover the Galactic Plane, the analysis excludes a region of the sky between $-10^{\circ} \leq b <10^{\circ}$ in galactic latitude. Furthermore, the regions around both celestial poles ($|\delta| <85^{\circ}$) are also excluded, as they were not included in the  7yr-PS search \cite{IceCube:7yrPS}. Finally, a cutoff of $z\leq 0.03$ is applied to the redshift of the objects from the 2MRS catalog, since it becomes magnitude-limited beyond that distance.\\

The degree of correlation between IceCube multiplets and baryon density (as mapped by the 2MRS catalog) is evaluated on a pixel-by-pixel product of all $N_{m}$ multiplets, using the following likelihood function:\\

\begin{equation}\label{multiplet_lh}
L(n_a) = \prod_{i=1}^{N_{m}}\left( \frac{n_a}{N_{m}} S_i + (1-\frac{n_a}{N_{m}})B_i \right)
\end{equation}\\

In this likelihood, the number of multiplets coming from the local universe $n_{a}$, normalized by the total number of selected multiplets $N_{m}$, is maximized for a given background ($B_i$) and signal ($S_i$) probability in all pixels $i$ containing a multiplet. The signal term $S_{i}$ of this likelihood is a normalized count of 2MRS objects contained in a HEALPix-defined pixel area of 3.36 square degrees. The full sky map of the signal PDF is shown in Fig. \ref{multiplet_signal}). The background term, $B_i$, is assumed to be constant across the map, and is thus the fraction of the sky covered by a pixel on the map, i.e. $A_{\mathrm{pixel}} /4\pi$.\\

The test statistic for the multiplet analysis is the same as Eq. \ref{eq:1}, the null hypothesis being in this case when $n_a=0$. The significance of the correlation is then estimated by comparing the value of the maximized likelihood ratio to the TS distribution calculated from background multiplet distributions randomized in RA, similar to the method used in the template analysis.

\section{Results}
\label{sec:results}
\subsection{Template Analysis}

\begin{figure} [htb]
\begin{center}
\includegraphics[width=.65\textwidth]{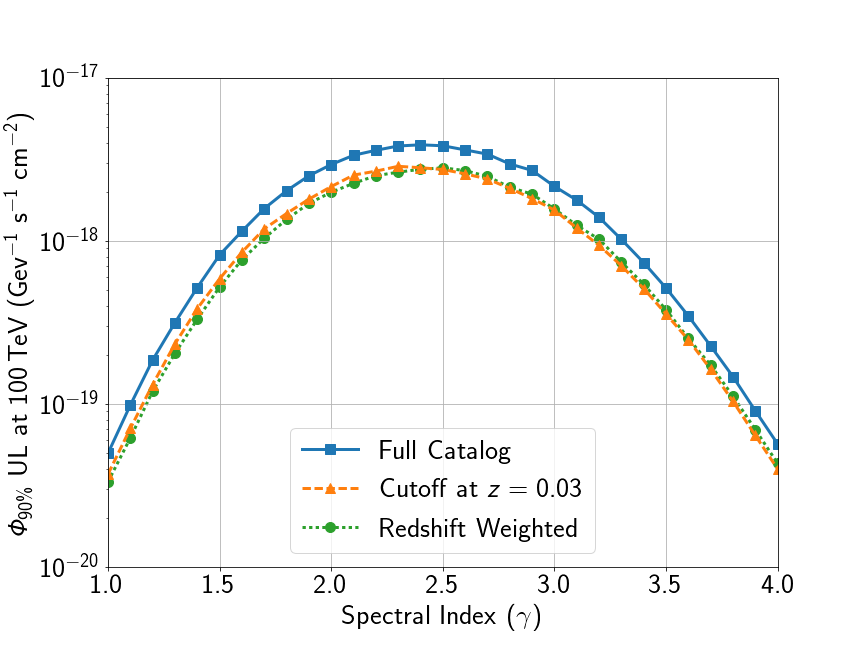} \\
\end{center}
\caption{Upper limit flux normalization for the \textit{template analysis} calculated for each of three source templates as a function of spectral index. Fluxes above these indicated limits are excluded by this analysis at 90\% confidence level. Flux is constrained due to the lack of excess events correlated with the spatial template and consistent with the energy range of the considered spectra.}
\label{fig:UL}
\end{figure}

All three correlation scenarios indicate an underfluctuation of neutrino events in the 7-yr PS sample. The data is therefore consistent with background-only null hypothesis.  Given the absence of a significant correlation, upper limits are calculated on the neutrino flux from local 2MRS galaxies in this catalog under these three hypotheses.  These limits are reported in Table \ref{table:results}. We note that if all galaxies in the Universe emit neutrinos equally, the galaxies in the 2MRS catalog would contribute only a small fraction of the total neutrino flux. The second column of the table indicates the TS value, while the third column gives the p-value or probability of obtaining an equal or larger TS assuming the background-only hypothesis.  The fourth column contains 90\% upper limits on the neutrino flux from local structure in the 2MRS catalog, assuming an $E^{-2}$ spectrum.  There is a systematic uncertainty on these upper limits of 11\% as identified in \cite{IceCube:7yrPS}.  By assuming a simple power law spectrum, spectral index $\gamma$ is scanned over and the 90\% upper limits are plotted as a function of $\gamma$ in Figure \ref{fig:UL}.  In column 5 of Table \ref{table:results}, the flux upper limits to an E$^{-2.19}$ spectrum are reported as a fraction of the diffuse astrophysical neutrino flux measured by IceCube \cite{Aartsen:2017mau}.  Physically, these values can be interpreted as 90\% upper limits on the astrophysical flux which originate from interactions of UHECRs with the local distribution of large-scale structure as seen in the 2MRS catalog.

\begin{table}[ht]
\begin{center}
\begin{tabular}{|c|c|c|c|c|} \hline 

Template & \begin{tabular}[x]{@{}c@{}}Test \\Statistic\end{tabular} &p-value & \begin{tabular}[x]{@{}c@{}}Upper Limit \\$\Phi_{90\%}$ Flux\end{tabular} & \begin{tabular}[x]{@{}c@{}}UL as Percentage of\\Measured Diffuse Flux\end{tabular}  \\
\hline
Full Catalog Template  & 0.0 & 1.0 & 2.89 $\times$ 10$^{-18}$ & 30\%\\
\hline
$z$ \textless 0.03 Template & 0.0 & 1.0 & 2.15 $\times$ 10$^{-18}$  & 22\%\\
\hline
\begin{tabular}[x]{@{}c@{}}Full Catalog Template  \\with redshift weighting\end{tabular}& 0.0 & 1.0  & 1.97 $\times$ 10$^{-18}$ & 20\% \\
\hline
\end{tabular} 
\end{center}
    \caption{Fluxes are integrated over the full sky and parameterized as dN/dE = $\Phi_{90\%}$ $\times$ ($\frac{E}{100\, \mathrm{TeV}}$)$^{-2}$ GeV$^{-1}$cm$^{-2}$s$^{-1}$ with 90\% confidence level upper limits.  The percentages shown in column 5 are based on IceCube measurements of the diffuse astrophysical flux of dN/dE = $1.01$ $\times$ $10^{-18}$ ($\frac{E}{100 \, \mathrm{TeV}}$)$^{-2.19}$ GeV$^{-1}$cm$^{-2}$s$^{-1}$sr$^{-1}$ \cite{Aartsen:2017mau}.}
\label{table:results}
\end{table}

\subsection{Multiplet Analysis}

The test statistic observed in the \textit{multiplet analysis} returned a value of 0.238, with a corresponding p-value of $80\%$, which is consistent with the null hypothesis of multiplets being uncorrelated with the 2MRS catalog.\\

\begin{figure}[ht]
    \centering
    \includegraphics[width=0.9\textwidth]{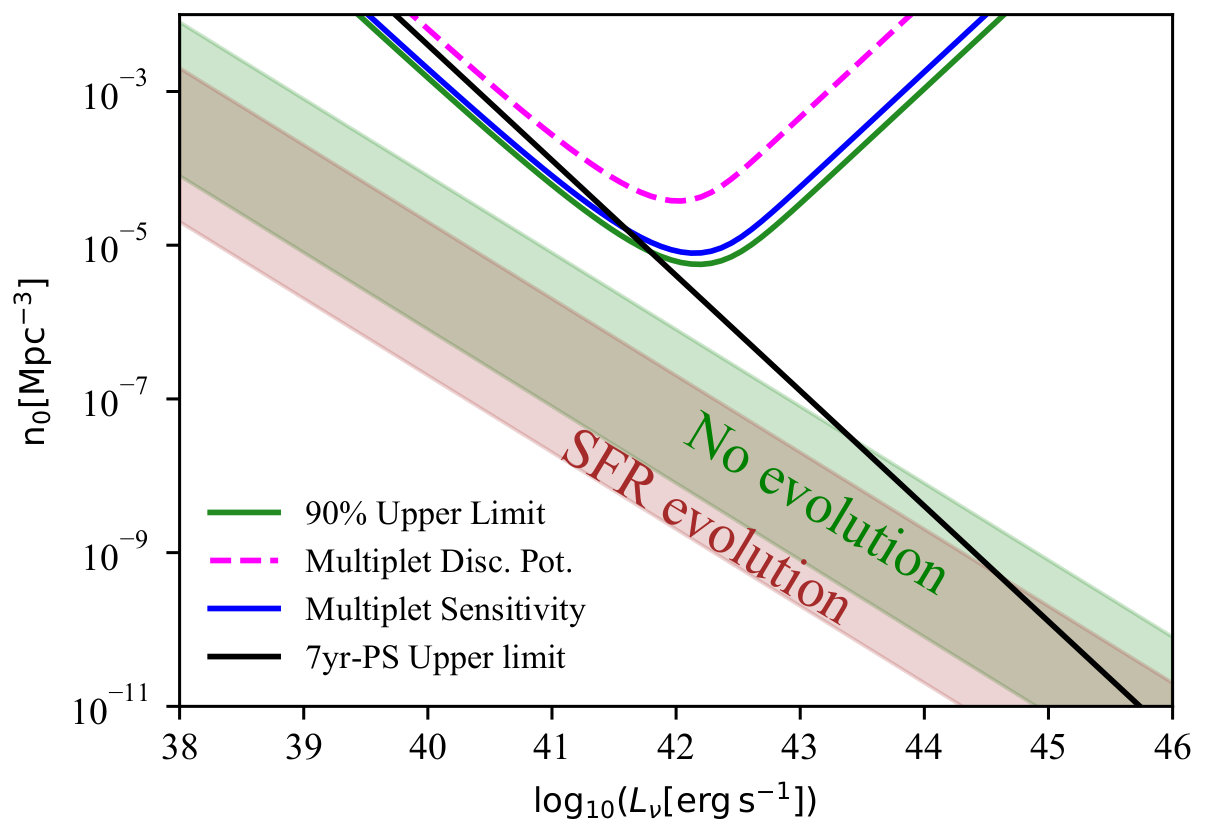}
    \caption{Limits on the density of neutrino source populations in the local universe, as a function of that population's neutrino luminosity $L_{\nu}$, assuming a uniform population of sources with an $E^{-2.2}$ spectrum. The observed limit from the multiplet analysis is shown by the green line, and can be compared to the 90\% sensitivity (blue) and discovery potential (pink) of the analysis method. The solid black line is a recalculation of the limit derived in \cite{Mertsch:2016hcd}, using a critical redshift matching the bound used in selecting 2MRS objects (z=0.03). The green band shows the parameter space covered if 2MRS objects were generating between 1\% to 100\% of the observed IceCube diffuse flux. The red band adds a correction factor, if one assumes a redshift evolution of neutrino sources that would match currently observed star formation rates (SFR).}
    \label{fig:kowalski}
\end{figure}

Given the absence of a detectable population of neutrino sources correlated with the selected multiplets, an upper limit is placed on the density $n_0$ and neutrino luminosity $L_{\nu}$ of the hypothesized population of sources. Figure \ref{fig:kowalski} shows the 90\% limit in the density-luminosity parameter space, derived using Eq. 2.17 of \cite{Mertsch:2016hcd}. This equation relates the best-fit value of $n_{a}$ obtained with the probability of seeing $n\geq 2$ neutrinos from sources distributed within a comoving volume reaching out to the boundary redshift of $z=0.03$. This probability can be analytically derived if one assumes that each source has a poisson probability of emitting neutrinos detectable by IceCube. The expectation value of this probability, $\lambda$, scales with the distance at which a source is able to emit 2 or more neutrinos detectable by IceCube, which itself scales with the assumed neutrino luminosity of the source, $L_{\nu}$. Integrating over the co-moving volume yields the total number of detectable multiplets, which is equivalent to multiplying a local density $n_0$ by an integrated redshift-dependent source density function.\\

The \textit{multiplet search} can be compared to the $n_0$-$L_\nu$ limit one would obtain from the non-observation of statistically significant neutrino clusters, which has been studied by \cite{Mertsch:2016hcd} on the 7yr-PS dataset, and by \cite{murase_waxman} in an earlier IceCube sample. The advantage of using a multiplet correlation technique can been seen in the case where the observed flux of neutrinos comes from a population of low-luminosity sources. For luminosities above a turnover point of $L_{\nu}\approx 10^{42}$ $\mathrm{erg}\cdot \mathrm{s}^{-1}$, sources at distances beyond $z=0.03$ are expected to emit neutrinos that are detectable by IceCube. Taking the conservative assumption that the universe becomes isotropic beyond that point, the multiplet analysis is then expected to become less sensitive, as correlations with the 2MRS catalog are expected to wash out.\\

Both the multiplet and 7yr-PS limits can be compared to the lower (higher) density of source population expected, if 1\% (100\%) of the diffuse flux observed by IceCube came from 2MRS objects. This range is illustrated by the shaded bands in Fig. \ref{fig:kowalski}. The green band assumes that the density of neutrino sources does not evolve with redshift, while the red band incorporates a correction factor to account for a more plausible evolution, as derived in \cite{Yuksel_SFR}. All elements in the plot assume a uniform population emitting neutrinos with an $E^{-2.2}$ spectrum.\\

\section{Conclusion}
\label{sec:conclusion}

Two searches for correlations between IceCube neutrinos and the 2MRS catalog were performed, probing two different neutrino emission scenarios. The \textit{template analysis} looked for excess neutrinos correlated with local galaxy column density. Meanwhile, the \textit{multiplet analysis} examined possible correlations between clusters of neutrinos and the same galactic matter distribution, which would be caused by the presence of sources with low neutrino luminosities within these local galaxies. None of these searches have observed significant correlations, and produce p-values that are entirely consistent with statistical background fluctuations.\\

Given a lack of correlation in the template analysis, an upper limit is placed on the flux from the galaxies in the 2MRS catalog, weighted according to three hypotheses.  These limits constrain the neutrino flux from a mostly isotropic diffuse UHECR flux interacting with baryonic matter traced by galaxies in the local universe. This analysis is limited by assumptions of the column density, distance, and completeness of the galaxies in the 2MRS survey, and may be improved with future redshift surveys.\\

Given a lack of correlations in the multiplet analysis, constraints were derived on the density of a hypothetical population of sources located within the local universe. These constraints are more stringent than the constraints obtained when one only looks for statistically significant neutrino clusters within the 7yr-PS dataset, in the case where the underlying population of sources has an average neutrino luminosity below $\approx 10^{42}$ erg$\cdot$s$^{-1}$.  Above this luminosity, one would expect neutrino multiplets to be visible from source beyond the redshift cutoff of $z=0.03$. Note that the multiplet analysis of 7 years of data is not yet constraining scenarios where 1\% to 100\% of the observed astrophysical neutrino flux comes from low-luminosity sources in the 2MRS catalog, in both redshift evolution scenarios (see Fig. \ref{fig:kowalski}).\\

The contribution of the local universe to the astrophysical neutrino flux in both searches could be significantly enhanced in scenarios in which the density of neutrino sources does not evolve (or evolve negatively) with redshift \cite{Taylor:2015rla}. The lack of a statistically significant excess from the local universe in either search thus serves as a constraint on scenarios with very strong negative redshift evolution of astrophysical neutrino sources.\\


\input{acknowledgements.tex}


\end{document}

%% file: authors_jcap.tex
\author[15]{M. G. Aartsen,}
\author[54]{M. Ackermann,}
\author[15]{J. Adams,}
\author[11]{J. A. Aguilar,}
\author[19]{M. Ahlers,}
\author[45]{M. Ahrens,}
\author[25]{C. Alispach,}
\author[36]{K. Andeen,}
\author[51]{T. Anderson,}
\author[11]{I. Ansseau,}
% ...
\affiliation[11]{Universit{\'e} Libre de Bruxelles, Science Faculty CP230, B-1050 Brussels, Belgium}
% ...
\affiliation[15]{Dept. of Physics and Astronomy, University of Canterbury, Private Bag 4800, Christchurch, New Zealand}
% ...
\affiliation[19]{Niels Bohr Institute, University of Copenhagen, DK-2100 Copenhagen, Denmark}
% ...
\affiliation[25]{D{\'e}partement de physique nucl{\'e}aire et corpusculaire, Universit{\'e} de Gen{\`e}ve, CH-1211 Gen{\`e}ve, Switzerland}
% ...
\affiliation[36]{Department of Physics, Marquette University, Milwaukee, WI, 53201, USA}
% ...
\affiliation[45]{Oskar Klein Centre and Dept. of Physics, Stockholm University, SE-10691 Stockholm, Sweden}
% ...
\affiliation[51]{Dept. of Physics, Pennsylvania State University, University Park, PA 16802, USA}
% ...
\affiliation[54]{DESY, D-15738 Zeuthen, Germany}
% ...

%% file: acknowledgements.tex
\section*{Acknowledgements}

The authors gratefully acknowledge the support from the following agencies and institutions: USA {\textendash} U.S. National Science Foundation-Office of Polar Programs,
U.S. National Science Foundation-Physics Division,
Wisconsin Alumni Research Foundation,
Center for High Throughput Computing (CHTC) at the University of Wisconsin-Madison,
Open Science Grid (OSG),
Extreme Science and Engineering Discovery Environment (XSEDE),
U.S. Department of Energy-National Energy Research Scientific Computing Center,
Particle astrophysics research computing center at the University of Maryland,
Institute for Cyber-Enabled Research at Michigan State University,
and Astroparticle physics computational facility at Marquette University;
Belgium {\textendash} Funds for Scientific Research (FRS-FNRS and FWO),
FWO Odysseus and Big Science programmes,
and Belgian Federal Science Policy Office (Belspo);
Germany {\textendash} Bundesministerium f{\"u}r Bildung und Forschung (BMBF),
Deutsche Forschungsgemeinschaft (DFG),
Helmholtz Alliance for Astroparticle Physics (HAP),
Initiative and Networking Fund of the Helmholtz Association,
Deutsches Elektronen Synchrotron (DESY),
and High Performance Computing cluster of the RWTH Aachen;
Sweden {\textendash} Swedish Research Council,
Swedish Polar Research Secretariat,
Swedish National Infrastructure for Computing (SNIC),
and Knut and Alice Wallenberg Foundation;
Australia {\textendash} Australian Research Council;
Canada {\textendash} Natural Sciences and Engineering Research Council of Canada,
Calcul Qu{\'e}bec, Compute Ontario, Canada Foundation for Innovation, WestGrid, and Compute Canada;
Denmark {\textendash} Villum Fonden, Danish National Research Foundation (DNRF), Carlsberg Foundation;
New Zealand {\textendash} Marsden Fund;
Japan {\textendash} Japan Society for Promotion of Science (JSPS)
and Institute for Global Prominent Research (IGPR) of Chiba University;
Korea {\textendash} National Research Foundation of Korea (NRF);
Switzerland {\textendash} Swiss National Science Foundation (SNSF).  The  IceCube  collaboration  acknowledges  the  significant  contributions  to  this  manuscript  from  Stephen Sclafani and Naoko Kurahashi Neilson.